\documentclass[12pt,a4paper]{article}
\usepackage{dcolumn}
\usepackage{bm}
\usepackage{amssymb}
\usepackage{tabularx}
\usepackage{booktabs}
\usepackage{amsmath}
\usepackage{subcaption}
\usepackage{diagbox} 
\usepackage{hyperref}
\usepackage{graphicx}
\usepackage{lmodern}
\usepackage[left=2cm,right=2cm,top=2cm,bottom=2cm]{geometry}
\usepackage{authblk}
\title{Genetic Algorithm based Inverse Potentials for Resonant States of $\alpha-^{12}C$ Using Variable Phase Approach}
\author[1]{Ayushi Awasthi}
\author[1]{Arushi Sharma}
\author[1]{Barbie}
\author[1]{Ishwar Kant}
\author[1]{O.S.K.S. Sastri}
\affil[1]{Department of Physics and Astronomical Sciences, Central University of Himachal Pradesh, Dharamsala, Bharat (India)}
\begin{document}
\maketitle
\begin{abstract}
Elastic scattering between $\alpha$-particles and $^{12}\mathrm{C}$ nuclei plays a crucial role in understanding resonance phenomena in light nuclear systems. In this work, we construct inverse potentials for resonant states in $\alpha$-$^{12}\mathrm{C}$ elastic scattering using the variable phase approach, in tandem with a genetic algorithm based optimization technique. The reference function for the potential in the phase equation is chosen as a combination of three smoothly joined Morse-type functions. The parameters of the reference function are genetically evolved to minimize the the mean squared error (MSE) between the numerically obtained scattering phase shifts and the expected values. The resulting inverse potentials accurately reproduce the resonance energies ($E_r$) and the resonance widths ($\Gamma_r$) for the $\ell^{\pi}$ states, $1^-$, $2^+$, $3^-$, and $4^+$, showing excellent agreement with experimental data. This computational approach to constructing inverse potentials serves as a complementary to conventional direct methods for investigating nuclear scattering phenomena.
 
\vspace{0.5cm}
\end{abstract} 

\section{Introduction}
The investigation of low-energy elastic scattering between alpha particles ($\alpha$) and carbon-12 ($^{12}C$) nuclei holds substantial significance in nuclear physics \cite{orlov2016asymptotic}. This system not only illuminates the nature of nuclear interactions between these particles but also provides critical insights into the structure of oxygen-16 ($^{16}O$) \cite{ando2016elastic}. Precise phase shift measurements derived from elastic $\alpha–^{12}C$ scattering experiments are instrumental in benchmarking and refining various theoretical approaches, such as effective Field Theory \cite{ando2016elastic,yoon2019effective}, R-matrix Theory \cite{angulo2000r}, and Asymptotic Normalization Coefficients (ANC) \cite{irgaziev2015resonance}. These models aim to comprehensively describe nuclear forces and the associated scattering phenomena \cite{ando2016elastic}. Due to the complexities of nuclear structure and dynamics, these scattering data are essential for constraining model parameters and improving the predictive power of nuclear interaction models.
These models strive to accurately capture the complex characteristics of nuclear forces in few-body systems. 
\\ An earlier and particularly influential investigation by Becker et al.~\cite{plaga1987scattering} set a standard in the field by providing a detailed phase shift analysis of $\alpha$--$^{12}\mathrm{C}$ elastic scattering using multilevel $R$-matrix formalism. Their measurements covered 35 angles in the laboratory range of $22^\circ$ to $163^\circ$ and 51 incident $\alpha$ energies between 1.0 and 6.6 MeV. By analyzing angular distributions with high precision, they extracted phase shifts for partial waves up to $\ell = 6$, and provided reduced $\alpha$-particle widths for the important subthreshold states at 6.92 MeV ($2^+$) and 7.12 MeV ($1^-$) in $^{16}$O- states known to dominate the low-energy behavior of the $^{12}\mathrm{C}(\alpha,\gamma)^{16}\mathrm{O}$ reaction. \\ 
A recent major advancement in this domain is the high-precision measurement of elastic $^{12}\mathrm{C}(\alpha, \alpha)^{12}\mathrm{C}$ scattering performed by Tischhauser et al.~\cite{tischhauser2009measurement}, which employed a large-angle detector array and covered a wide energy range of 2.6–8.2 MeV. This comprehensive dataset allowed for a detailed $R$-matrix analysis \cite{anghel2022r} and the extraction of phase shifts using a novel Monte Carlo method, resulting in significantly improved statistical confidence compared to earlier studies. These phase shifts provide essential inputs for fine-tuning theoretical models that seek to describe not only scattering processes but also resonant behavior in $^{16}\mathrm{O}$. 
Beyond its nuclear structure relevance, elastic $\alpha–^{12}C$ scattering plays a pivotal role in nuclear astrophysics, particularly in the context of the $^{12}C(\alpha,\gamma)^{16}O$ radiative capture reaction \cite{brune2015radiative}. This reaction, in which an alpha particle is captured by a carbon-12 nucleus to form oxygen-16, is a cornerstone process in stellar helium burning \cite{deboer2017c}. The resulting carbon-to-oxygen ratio is a key factor in stellar evolution and the synthesis of heavier elements. However, direct measurement of the astrophysical S-factor at the Gamow peak energy (~0.3 MeV), which is relevant for stellar environments, is hindered by the Coulomb barrier \cite{ando2016elastic,yoon2019effective}. Consequently, theoretical models must extrapolate the cross section to low energies using experimental data obtained at higher energies \cite{ando2016elastic,yoon2019effective}. In this context, accurate elastic scattering data at low energies serve as a vital constraint for these extrapolations, enhancing the reliability of stellar reaction rate predictions. Additionally, such studies contribute to the determination of energy levels in oxygen isotopes, further underscoring the broad impact of this research area \cite{angulo2000r}.\\ 
An alternative and insightful method frequently employed in the literature for calculating scattering phase shifts is the \textit{Variable Phase Approach} (VPA), also known as the \textit{Phase Function Method} (PFM) \cite{calogero1967variable}. Unlike traditional techniques that require the explicit computation of wavefunctions \cite{mackintosh2012inverse,alhaidari2008j}, the VPA determines phase shifts by solving a first-order nonlinear Riccati-type differential equation  that depends directly on the interaction potential $V(r)$ \cite{babikov1967phase,zhaba2016phase}. A particularly powerful feature of the variable phase approach is its inherent inverse capability. That is, given experimental phase shift data as input, the method can be employed to construct the corresponding interaction potential $V(r)$. This inverse application makes the VPA a valuable tool not only for forward scattering calculations but also for deducing effective potentials from scattering observables \cite{khachi2023deuteron}. 
\\ The main challenge in this approach lies in accurately modeling the interaction potential \( V(r) \). In previous studies, \( V(r) \) has been expressed as a combination of nuclear and Coulomb potentials. For the nuclear part, potentials such as the Morse \cite{khachi2023inverse} and Malfliet Tjon \cite{awasthi2022s} forms have been employed, while for the Coulomb interaction, various ansätze like the error function and the atomic Hulthén potential have been considered~\cite{awasthi2024elastic}. 
However, these functional forms present a significant drawback, they do not satisfy Taylor's criteria~\cite{taylor1974new}, which are essential for a proper treatment of the Coulomb interaction in scattering problems. Therefore, there is a need to construct \( V(r) \) in a manner that inherently captures the full interaction-both nuclear and Coulomb-without relying on predefined forms for each. \\ To address this, we propose modeling \( V(r) \) as a purely mathematical function composed of three Morse-type potentials, smoothly joined at intermediate boundaries \cite{selgmorse}. This formulation allows for greater flexibility and adherence to the physical constraints required for scattering studies, while avoiding the limitations of conventional ansätze. 
Recently, our research group has successfully utilized this approach to construct inverse potentials for a two-body scattering systems, including $n-p$, $\alpha$-$\alpha$, $\alpha$-$^3$He,  $\alpha$-$^3$H, and $\alpha$-d systems \cite{np,alpha_alpha,kant2024ab,sharma2025machine}. These studies underscore the versatility and robustness of the VPA in both direct and inverse scattering problems across a wide range of nuclear interactions. In this study, we extend the methodology to construct inverse potentials corresponding to the resonant states with angular momenta \( \ell^{\pi} = 1^{-}, 2^{+}, 3^{-}, \) and \( 4^{+} \) of the 
$\alpha-^{12}C$ system. Using these potentials, we obtained the scattering phase shifts and subsequently calculated the corresponding resonance energies and resonance widths for each state.
\section{Methodology}
In this section, we outline the methodology employed to construct the interaction potential for the $\alpha-^{12}C$ scattering system. We begin by discussing the Variable Phase Approach (VPA), followed by a detailed description of the modeling strategy for the interaction potential \( V(r) \). Finally, we present the optimization algorithm used to determine the model parameters that best characterize the potential, which is physically relevant.
\subsection{Variable Phase Approach (VPA)}
The forward problem in nuclear scattering involves solving the three-dimensional, time-independent Schrödinger equation with a given potential \( V(r) \). This yields phase shifts, which characterize the scattering process and are extracted from experimental data by fitting partial waves to measured cross sections~\cite{hans2008nuclear}.
\\ However, VPA or phase function method offers an effective approach to the inverse scattering problem by reformulating the Schrödinger equation as a first-order Riccati equation~\cite{calogero1967variable,palov2021vpa}. The resulting phase function captures the accumulated phase shift due to a truncated potential, providing a clear distinction from free-particle behavior~\cite{babikov1967phase,balassa2023fixed}.
\\For elastic scattering by a central or arbitrary potential, the second-order Schrödinger equation can be reformulated as a Riccati-type equation~\cite{calogero1967variable,viterbo2014variable}:
\begin{equation}
\frac{d\delta_\ell(k,r)}{dr} = -\frac{U(r)}{k} \left[\cos\delta_\ell(k,r)\,\hat{j}_\ell(kr) - \sin\delta_\ell(k,r)\,\hat{\eta}_\ell(kr)\right]^2,
\label{PFMeqn}
\end{equation}
where \( k = \sqrt{2\mu E_{cm}/\hbar^2} \), \( U(r) = 2\mu V(r)/\hbar^2 \), and \( \mu \) is the reduced mass. The center-of-mass energy is given by \( E_{cm} = \frac{m_T}{m_T + m_P} E_{\ell ab} \), with \( m_T \) and \( m_P \) as the target and projectile masses, respectively. The phase function is initialized with \( \delta_\ell(0) = 0 \), and its asymptotic value \( \delta_\ell(\infty) \) yields the total phase shift. The Riccati-Hankel function relates to the spherical Riccati-Bessel and Neumann functions as \( \hat{h}_\ell(r) = -\hat{\eta}_\ell(r) + i\,\hat{j}_\ell(r) \). For $\ell$ $\neq$ 0, the Ricatti-Bessel and Riccati-Neumann functions used in VPA can be easily obtained by using following recurrence formulas \cite{calogero1967variable}:
\begin{equation}
   {\hat{j}_{\ell+1}}(kr)=\frac{2\ell+1}{kr} \hat{j_\ell}(kr)-{\hat{j}_{\ell-1}}(kr),
    \label{R1}
\end{equation}
\begin{equation}
   {\hat{\eta}_{\ell+1}}(kr)=\frac{2\ell+1}{kr} \hat{\eta_\ell}(kr)-{\hat{\eta}_{\ell-1}}(kr),
   \label{R2}
\end{equation}
Using these recurrence relations along with Eq.~\ref{PFMeqn}, the phase equation can be extended to higher \( \ell \) channels. Eq. \ref{PFMeqn} is a nonlinear differential equation, which we solve numerically using the fifth-order Runge-Kutta (RK5) method with the initial condition \( \delta_{\ell}(k,0) = 0 \). Since the potential enters the phase equation as a multiplicative term, the resulting potential profile may exhibit irregularities. To ensure smoothness and stability, we employ a reference potential approach, initializing \( V(r) \) with well-defined mathematical functions.
\subsection{Reference Potential Approach}
To model the interaction potential, we adopt the Reference Potential Approach (RPA), where \( V(r) \) is chosen as a combination of smoothly joined Morse-type functions \cite{selg2006reference}:
\begin{equation}
V_i^{\text{RPA}}(r) = V_i \pm D_i \left[e^{-2\alpha_i(r - r_i)} - 2e^{-\alpha_i(r - r_i)}\right], \quad i = 1, 2, \ldots
\label{U}
\end{equation}
Here, \( D_i \) denotes the depth, \( r_i \) the equilibrium position, and \( \alpha_i \) shape of the function. These functions are smoothly joined at intervening points \( x_{i+1} \). While additional components improve the fit to experimental data, they also increase the model's complexity~\cite{selg2016complete}.
Recently, Awasthi et al.~\cite{np} used three Morse components (\(i = 0, 1, 2\)) to construct high-precision inverse potentials for neutron-proton (n-p) scattering, capturing interactions across all three (short range, intermediate and long range) regions. As the $\alpha-^{12}C$ scattering is a charge particle scattering, so to model the interaction for this system, we need to invert the third component of the Morse function , so that it captures the Coulomb interaction explicitly without using any existed anstazes. The reference potential \( V(r) \) for $\alpha-^{12}C$ scattering is given by:
\begin{eqnarray*}
V(r) =
\begin{cases} 
V_{R}(r) = V_0 + D_0 \left( e^{-2 \alpha_0 (r - r_0)} - 2 e^{-\alpha_0 (r - r_0)} \right) & \text{if } r \leq x_1 \\
V_{NC}(r) = V_1 + D_1 \left( e^{-2 \alpha_1 (r - r_1)} - 2 e^{-\alpha_1 (r - r_1)} \right) & \text{if } x_1 < r < x_2 \\
V_{CL}(r) = V_2 - D_2 \left( e^{-2 \alpha_2 (r - r_2)} - 2 e^{-\alpha_2 (r - r_2)} \right) & \text{if } r \geq x_2
\end{cases}
\end{eqnarray*}
where \( V_{R} \), \( V_{NC} \), and \( V_{CL} \) represent the short-range nuclear, short-range nuclear-Coulomb, and long-range Coulomb interactions, respectively. To ensure smoothness at the boundary points \( x_1 \) and \( x_2 \), both the potential and its derivatives must be continuous:
\begin{eqnarray*}
V_{R}(r) \big|_{x_1} = V_{NC}(r) \big|_{x_1}, \quad V_{NC}(r) \big|_{x_2} = V_{CL}(r) \big|_{x_2}, \\
\frac{dV_{R}(r)}{dr} \big|_{x_1} = \frac{dV_{NC}(r)}{dr} \big|_{x_1}, \quad \frac{dV_{NC}(r)}{dr} \big|_{x_2} = \frac{dV_{CL}(r)}{dr} \big|_{x_2}.
\end{eqnarray*}
With four parameters for each Morse component and two boundary points, the reference potential comprises a total of 14 parameters. Imposing four continuity conditions at the intervening boundaries reduces the number of independent parameters to 10. Consequently, the model is characterized by 8 model parameters along with the boundary points \( x_1 \) and \( x_2 \). To construct the interaction potential, these 10 parameters are optimized using a machine learning-based algorithm.
\subsection{Genetic Algorithm-Based Optimization}
\begin{figure}
    \centering
    \includegraphics[scale=0.5]{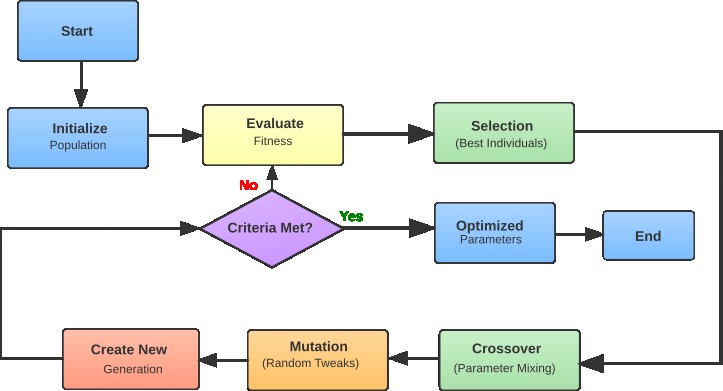}
    \caption{Flowchart for the optimization of model parameters using a genetic algorithm.}
    \label{flow}
\end{figure}

Advanced computational methods, particularly machine learning (ML) algorithms, have proven effective for parameter optimization in quantum systems \cite{koziel2011computational}. These approaches efficiently explore high-dimensional spaces and minimize discrepancies between theory and experiment, enabling accurate construction of interaction potentials \cite{wang2024machine}. Among them, genetic algorithms (GAs) are especially suited for complex, non-linear optimization tasks \cite{man1996genetic,mccall2005genetic,katoch2021review}, leveraging evolutionary strategies to adjust multiple parameters simultaneously.
In this study, we have used GA to optimized the model parameters of the reference function. GAs are stochastic search heuristics inspired by the principles of natural selection and genetics. The algorithm begins with a randomly generated population of candidate solutions, each representing a unique set of model parameters. Through iterative application of genetic operators-selection, crossover, and mutation-the population evolves toward an optimal solution. The fitness of each candidate is evaluated by comparing the theoretical phase shifts obtained from the reference potential to expected data. This population-based approach allows GAs to effectively navigate rugged and high-dimensional error landscapes, making them ideal for the global optimization required in inverse scattering problems \cite{thompson2025gaor}. 
The flow chart for optimization of model parameters is shown in Fig \ref{flow}.
The optimization begins with the initialization of the reference potential function, \( V(r) \), which is constructed using a combination of smoothly joined Morse functions. Initially, a set of candidate parameters is chosen randomly within a defined search space, ensuring the exploration of various potential solutions. The next step involves evaluating the initial population by calculating the error between the theoretical phase shifts predicted by the reference potential and the experimental data. This error is quantified using a cost or fitness function, which measures the deviation between the model’s predictions and the observed experimental values. In our work, we use the Mean Squared Error (MSE) as the loss function, defined as:
\begin{equation}
    MSE= \frac{1}{N}\sum_{i=1}^N \left(\delta_i^{exp}-\delta_i^{sim}\right)^{2}
\end{equation}
\\Using the fitness values computed from the initial population, the GA proceeds by selecting individuals with lower error values, mimicking the process of natural selection. These individuals are subjected to genetic operations, such as crossover and mutation, which generate new candidate solutions. The crossover operation combines the parameters of two parent solutions to create offspring with potentially better performance, while mutation introduces random changes to the parameters to explore new areas of the parameter space. This process of selection, crossover, and mutation is repeated over multiple generations, with each generation potentially providing improved parameter sets. At each step, the fitness of the new population is recalculated, and the selection process ensures that only the most optimal parameter sets are passed on to the next generation. The algorithm terminates once a predefined convergence criterion is met, typically when the error between the predicted and experimental data reaches an acceptable minimum.
\section{Results and Discussion}
\subsection{Database:}
\begin{figure}[h!]
    \centering
    \includegraphics[scale=0.8]{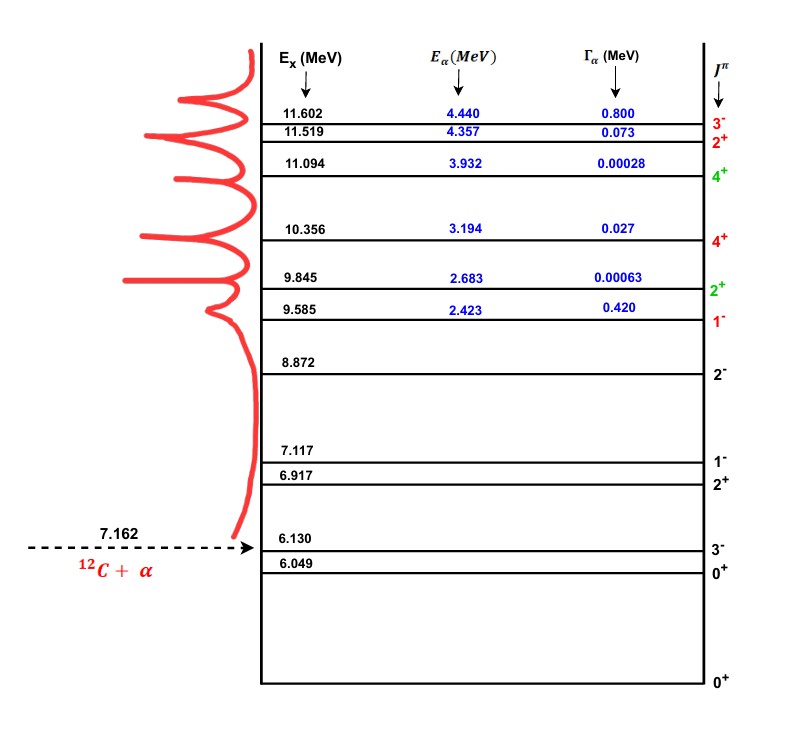}
    \caption{Energy Level Diagram of $^{16}O$ above the $\alpha-^{12}C$ threshold.}
    \label{energy}
\end{figure}

The phase shift data for various partial waves (\( \ell^{\pi} = 1^{-} \) to \( \ell^{\pi} = 4^{+} \)) have been taken from Plaga et al.~\cite{plaga1987scattering}. Although data are available for a wide range of $\ell$ channels, our study focuses on the resonant states of the \(\alpha\text{--}^{12}\mathrm{C}\) system corresponding to $\ell^{\pi}$ = \(1^-\), \(2^+\), \(3^-\), and \(4^+\). 
The energy level scheme of $^{16}O$ (above the \(\alpha\text{-}^{12}\mathrm{C}\) threshold ) reveals multiple resonances within the same spin-parity configurations \cite{tilley1993energy} as shown in Fig. \ref{energy}. The red peaks indicate the resonance states of the corresponding levels.
In particular, both the \(2^+\) and \(4^+\) states exhibit two resonances, one sharp (narrow) and one broad. These features suggest a more complex underlying nuclear structure and scattering dynamics. In this work, we focus on constructing inverse potentials corresponding to the broader resonances ($\ell^{\pi} = 1^{-}, 2^{+}, 3^{-}$) and the narrow resonance ($\ell^{\pi} = 4^{+}$), as these are more reliably represented in the experimental phase shift data. To effectively capture the resonance behavior, the data curation was performed with careful consideration of the underlying physical features~\cite{awasthi2024elastic}. The energy range up to 6.56~MeV was partitioned into three distinct regions based on observed trends in the phase shift data. For the partial wave \(\ell^{\pi} = 1^{-}\), the energy intervals were defined as 1.2--2.2~MeV, 2.2--3.7~MeV, and 3.8--4.5~MeV. Similarly, for \(\ell^{\pi} = 2^{+}, 3^{-},\) and \(4^{+}\), the regions were chosen as 2--4.5~MeV, 4.5--6.11~MeV, and 6.11--6.56~MeV. This classification was guided by a detailed analysis of the energy dependence of the phase shifts. In both cases, the phase shifts exhibit minimal variation in the first and third regions, while the second region shows a pronounced increase in the derivative \( \frac{d\delta(E)}{dE} \), indicative of resonant behavior in the corresponding $\ell$ states.
\subsection{Optimization of  Inverse Potential for $\alpha-^{12}C$ scattering using Genetic Algorithm}
\begin{figure*}[htbp]
\centering
\begin{subfigure}{0.45\linewidth}
\centering
\includegraphics[scale=0.35]{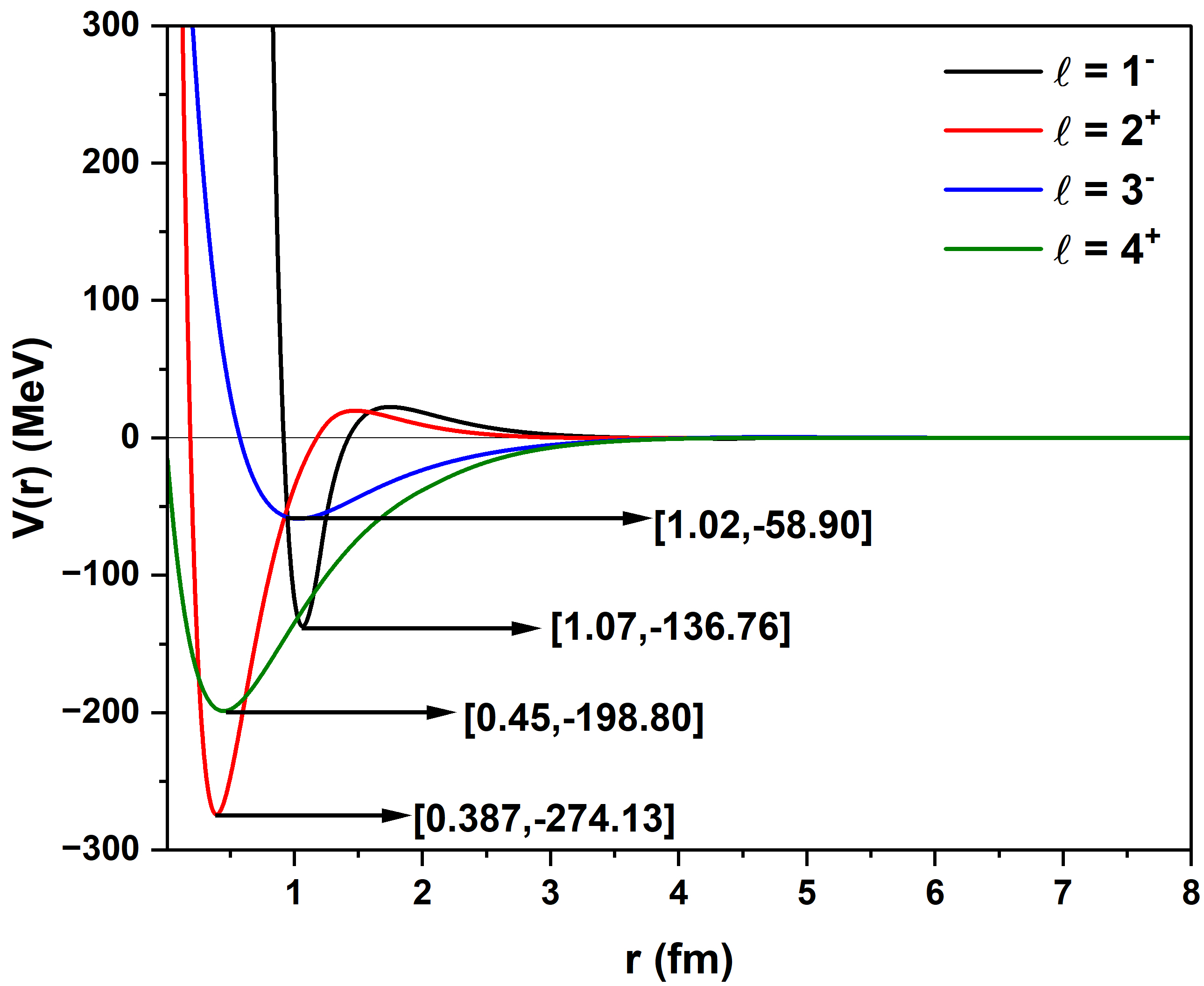}
\caption{Optimized inverse potentials of  elastic $\alpha$–$^{12}\text{C}$ scattering for $\ell^{\pi} = 1^-, 2^+, 3^-$, and $4^+$ partial waves.}
\label{pot}
\end{subfigure}
\hspace{0.5cm} 
\begin{subfigure}{0.45\linewidth}
\centering
\includegraphics[scale=0.35]{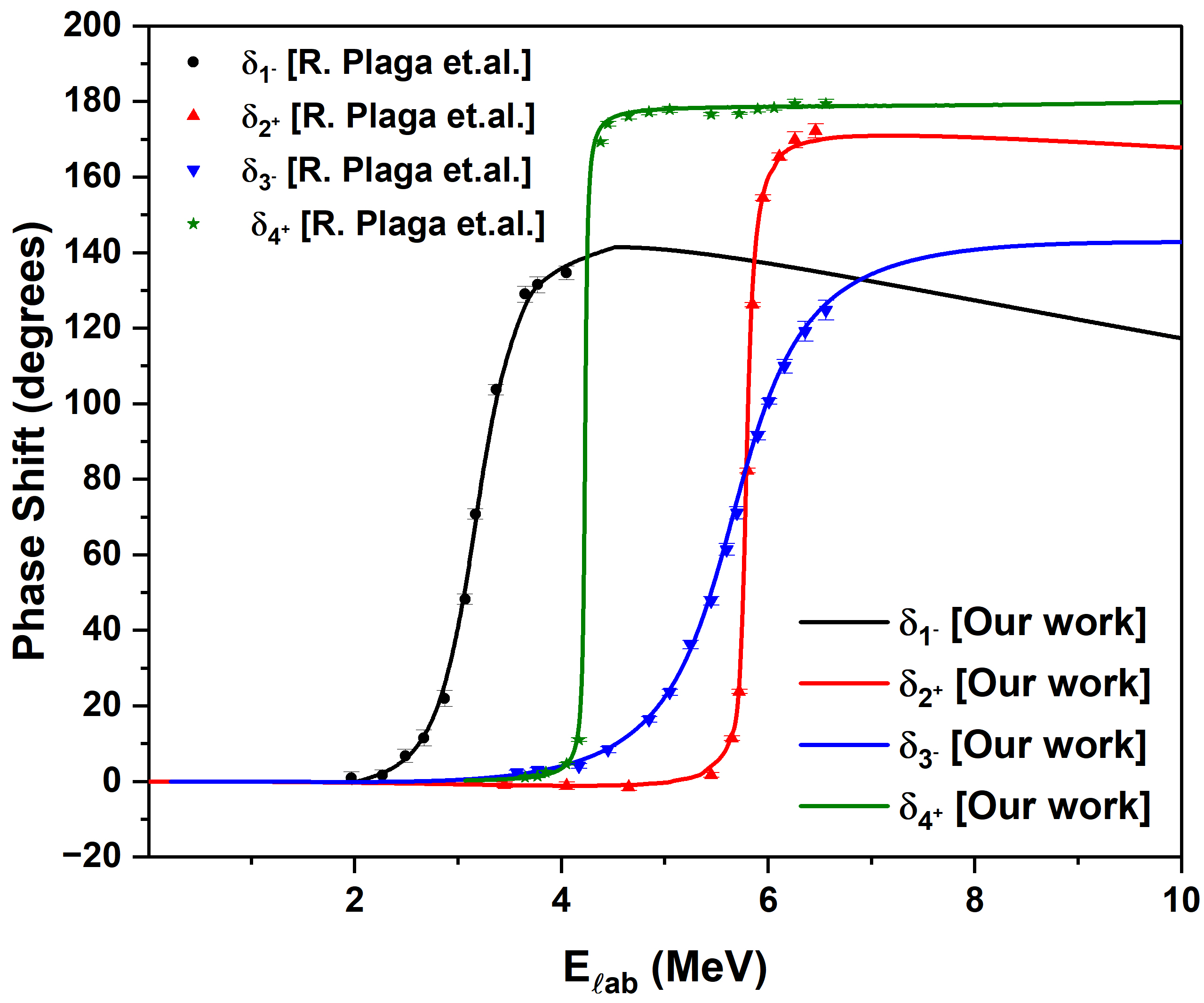}
    \caption{Simulated Scattering Phase shifts along with expected scattering phase shifts \cite{plaga1987scattering} }
    \label{sps}
\end{subfigure}
\end{figure*}

By utilizing the curated datasets as input within the framework of the variable phase approach, we optimized the model parameters of the chosen reference functions and construct the inverse potentials for the resonant states of the $\alpha$–$^{12}\text{C}$ system. The reference potential is defined by 10 independent parameters, leading to a 10-dimensional parameter space for optimization. A genetic algorithm (GA) was employed to perform this optimization, leveraging evolutionary strategies such as selection, crossover, and mutation to efficiently explore the solution space. The performance of the optimization process is highly sensitive to the choice of parameter bounds. The final integration limit, $r_f$, was selected to be sufficiently large to ensure that the potential asymptotically vanishes at long distances. The initial population of candidate solutions was randomly generated within the defined bounds using a fixed random seed. While this broad sampling enhances diversity in the trial potentials, it can also lead to an increase in the mean square error (MSE) and a longer convergence time.
Following 10,000 iterations and detailed analysis of the resulting phase shifts and potential profiles, the parameter space was systematically refined. This refinement reduced the range of sampled values, leading to a decrease in computational time and improved phase shift accuracy. A key advantage of the genetic algorithm lies in its ability to explore a broad spectrum of potential shapes while converging reliably toward an optimal solution. The final optimized model parameters for each $\ell$-channel are presented in Table~\ref{model}. It is noteworthy that the parameter $V_2$ consistently approaches zero (on the order of $10^{-5}$) across all channels. Therefore, it is omitted from the table for clarity. The minimum MSE values corresponding to the best-fit potentials are highlighted in bold in Table~\ref{model}.
\begin{table}[htbp]
\caption{Optimized model parameters \textbf{(P)} of the reference function for different states \textbf{($\ell^{\pi}$)} along with the corresponding Mean Square Error (MSE) using a genetic algorithm.}
\label{model}
\begin{tabularx}{\textwidth}{|c|*{9}{>{\centering\arraybackslash}X}|c|}
\hline
\textbf{$\ell^{\pi}$}/\textbf{P} 
& $\alpha_0$ ($\text{fm}^{-1}$) & $\alpha_1$ ($\text{fm}^{-1}$) & $\alpha_2$ ($\text{fm}^{-1}$)
& $r_0$ (fm) & $r_1$ (fm) & $r_2$ (fm) 
& $x_1$ (fm) & $x_2$ (fm) & $D_0$ (MeV)
& \textbf{MSE} \\ \hline
$1^{-}$ & 2.89 & 2.11 & 8.19 & 1.06 & 1.75 & 4.41 & 1.21 & 4.40 & 483.28 & \textbf{1.87} \\ \hline
$2^{+}$ & 3.19 & 2.33 & 4.42 & 0.39 & 1.47 & 5.76 & 1.22 & 2.94 & 325.64 & \textbf{1.51} \\ \hline
$3^{-}$ & 1.47 & 1.40 & 2.14 & 1.03 & 0.80 & 4.72 & 1.55 & 4.60 & 63.46 & \textbf{0.75} \\ \hline
$4^{+}$ & 1.55 & 1.82 & 3.28 & 0.45 & 1.57 & 2.93 & 2.00 & 3.93 & 194.23 & \textbf{0.89} \\ \hline
\end{tabularx}
\end{table}
Using the optimized model parameters, we construct the inverse potential and compute the corresponding scattering phase shifts by solving the phase equation. These results are presented in Fig. \ref{pot} and Fig. \ref{sps}, respectively.
From these figures, following observations are made:
\begin{enumerate}
\item For \( \ell^{\pi} = 1^{-} \), the potential comes to be deeper with a depth of \( -198.80 \, \text{MeV} \), with the maximum attraction occurring at \( r = 1.07 \, \text{fm} \). The trend of the phase shift data reveals a consistent increase, albeit in a nonlinear fashion. This suggests that the potential should possess both repulsive and attractive components. Consequently, the resulting potential exhibits a repulsive core for \( r > 1 \, \text{fm} \), an attractive region up to \( r = 1.5 \, \text{fm} \), and a Coulomb barrier of \( 22.43 \, \text{MeV} \) at \( r = 1.74 \, \text{fm} \), and  after 3 fm, potential approaches to zero.
\item For \( \ell^{\pi} = 2^+ \), the potential becomes significantly deeper than for \( l = 1^- \), with a depth of \( -274.13 \, \text{MeV} \) at \( r = 0.38 \, \text{fm} \). The phase shift data increases steadily up to \( 4.5 \, \text{MeV} \), after which there is a sharp increase to \( 6.2 \, \text{MeV} \), followed by a continuous rise until it reaches \( 6.56 \, \text{MeV} \). This behavior suggests that the potential should be predominantly attractive, while also possessing repulsive components. The resulting potential exhibits a hard-core repulsion for \( r < 0.35 \, \text{fm} \), followed by a strong attraction up to \( r = 1.3 \, \text{fm} \), and a Coulomb barrier of \( 19.63 \, \text{MeV} \) at \( r = 1.44 \, \text{fm} \). Beyond \( r = 3 \, \text{fm} \), the potential approaches to zero.
\item For \( \ell^{\pi} = 3^- \), the potential is not as deep compared to \( l = 1^- \) and \( l = 2^+ \). The depth of the potential is \( -58.90 \, \text{MeV} \) at \( r = 1.02 \, \text{fm} \). Based on the trend of the phase shift data, it can be observed that the potential increases consistently, but in a nonlinear manner, suggesting that the potential contains both attractive and repulsive components. The obtained inverse potential shows repulsion up to \( r = 0.57 \, \text{fm} \), followed by attraction up to \( r = 3.445 \, \text{fm} \). Beyond this, the Coulomb potential appears, with a barrier of \( 0.5 \, \text{MeV} \) at \( r = 4.72 \, \text{fm} \).
\item  For $\ell^{\pi} = 4^+$, the depth of the potential reaches $-137.70$ MeV at $r = 1.07$ fm, which is deeper than that for $l = 3^-$. The corresponding scattering phase shifts exhibit a sharp rise up to nearly $180^\circ$, indicating a strong attractive interaction and the presence of a shape resonance.
\end{enumerate}
From these plots, it is evident that the Coulomb barrier emerges naturally, without explicitly imposing any specific ansatz for the Coulomb interaction. Thus, our methodology captures the entire effective interaction between the nuclei without requiring an explicit description of the internal interaction of the constituent nucleons.
\subsection{Effect of the Centrifugal Term on the Post-Optimized Inverse Potential}
\begin{figure}[h!]
\centering
\includegraphics[scale=0.4]{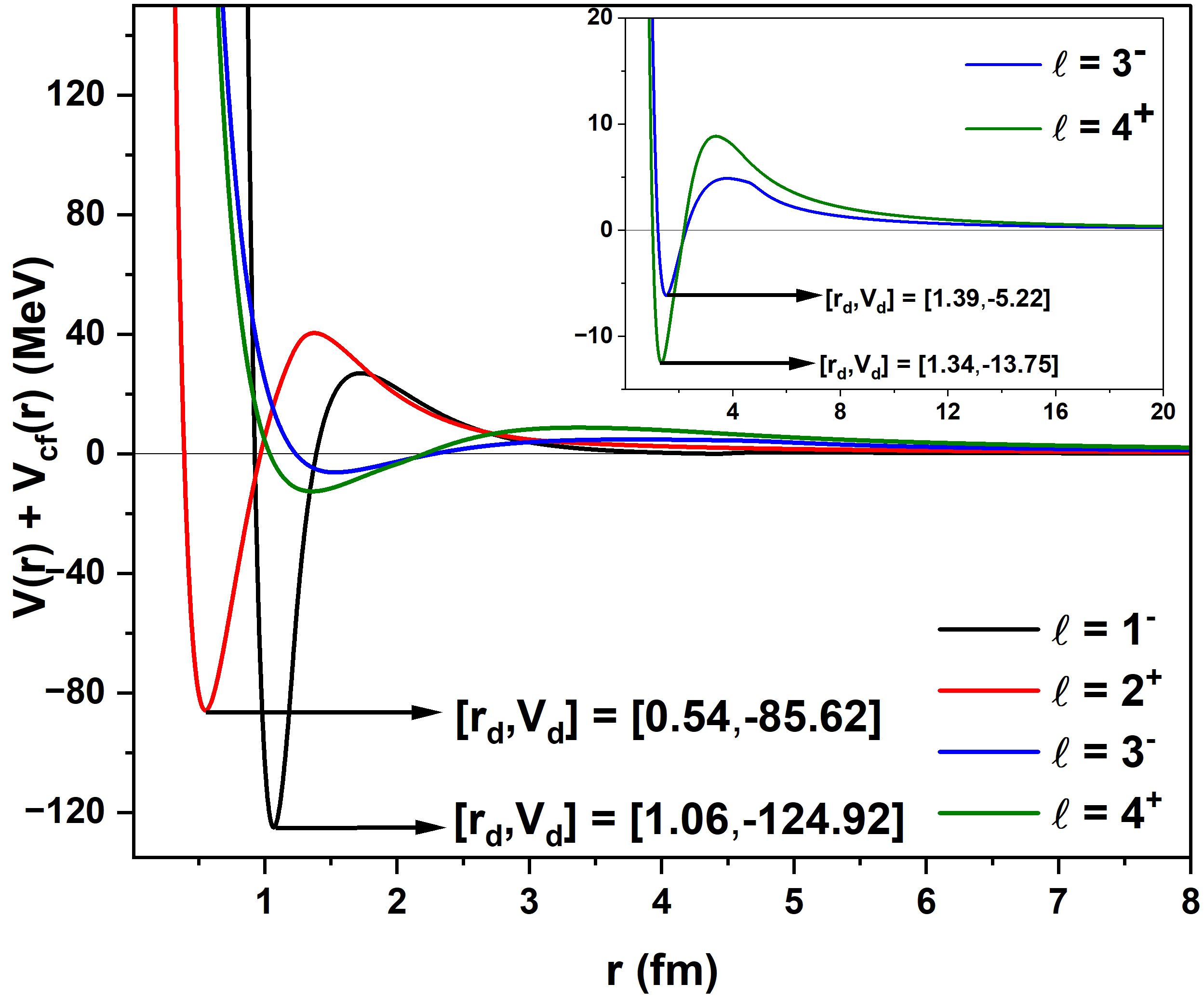}
\caption{Inverse Potentials after adding centrifugal term for different $\ell^{\pi}$ channels.}
\label{Centri}
\end{figure}

After obtaining the inverse potential, the influence of the centrifugal term on the resulting interaction becomes evident. The energy level diagram of ${}^{16}\text{O}$ exhibits the sequence of levels as $1^-$, $2^+$, $4^+$, and $3^-$. Consequently, the depth of the potential should reflect this ordering. 
\\ From Fig.~\ref{Centri}, it can be observed that, after including the centrifugal term, the sequence of the potential depths aligns with the expected order derived from the energy level diagram. Specifically, the potentials associated with the $1^-$ and $2^+$ states exhibit greater depths compared to those corresponding to the $4^+$ and $3^-$ states, in agreement with the structure of the energy spectrum.  The key characteristics of these potentials are their depth \( V_d \) at the position \( r_d \), which indicates the point of maximum attraction, and the Coulomb barrier height \( V_{CB} \) at \( r_{CB} \). These are referred to as the interaction parameters. It is important to note that the optimized model parameters do not directly represent the physical nature of the interaction. Instead, the interaction parameters \( V_d \) and \( V_{CB} \) are obtained from the total potential, which includes the centrifugal term, evaluated at \( r_d \) and \( r_{CB} \), respectively. The calculated interaction parameters for different \( (\ell^{\pi}) \) channels are shown in Table~\ref{param}. 
\begin{table}[htbp]
\centering
\caption{Interaction parameters for different $\ell^{\pi}$ channels of $\alpha$–$^{12}$C scattering after inclusion of the centrifugal term.}
\begin{tabular}{|c|c|c|c|c|}
\hline
~~$\ell^{\pi}$~~  & ~~$r_d$ (fm)~~ &~~ $V_d$ (MeV)~~ & ~~$r_{CB}$ (fm)~~ & ~~$V_{CB}$ (MeV)~~ \\ \hline
$1^-$ & 1.06    & -124.92  & 1.77     & 27.59     \\ \hline
$2^+$ & 0.54    & -85.62   & 1.38     & 40.46     \\ \hline
$3^-$ & 1.39    & -5.22    & 3.84     & 4.85      \\ \hline
$4^+$ & 1.34    & -13.75   & 3.26     & 8.8       \\ \hline
\end{tabular}
\label{param}
\end{table}
From table~\ref{param}, one can observe that the ordering of the potential depths \( V_d \) is consistent with the expected energy level hierarchy of the system. This consistency indicates that the constructed interaction potentials are physically meaningful and capable of reproducing key nuclear properties, such as the resonance energies and resonance widths of the system. Thus, the methodology not only successfully constructs the interaction potentials but also provides physically relevant descriptions that are in agreement with the known nuclear structure.
\subsection{Partial Cross Section and Resonance Parameters}
\begin{figure}[h!]
\centering
\includegraphics[scale=0.35]{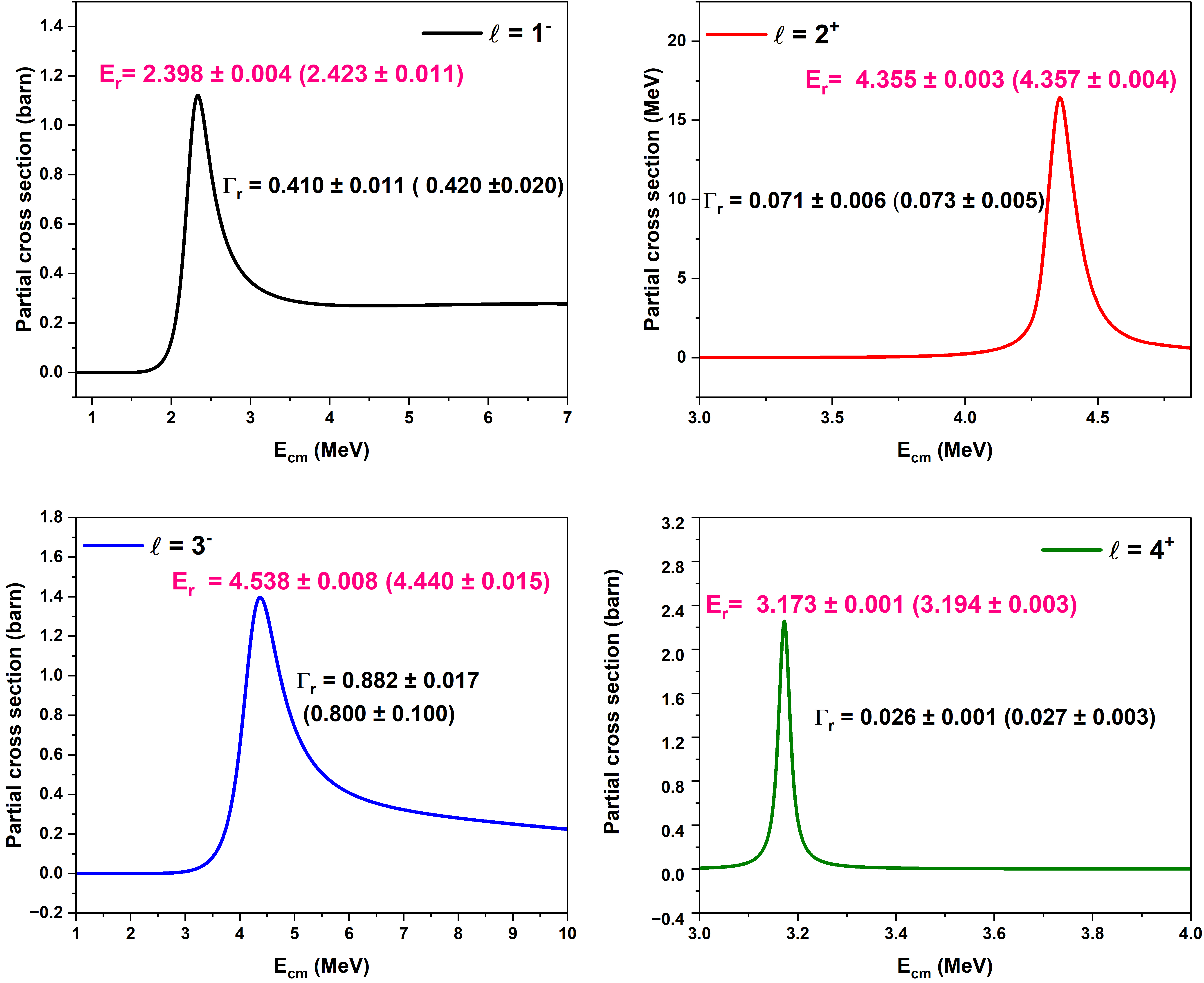}
\caption{Obtained Partial cross sections for different resonant states of $\alpha-^{12}C$ along with calculated resonance energies $E_{r}$ and resonance width $\Gamma_{r}$ in MeV. The experimental values \cite{epaps_12C_alpha} of $E_{r}$ and $\Gamma_{r}$ are given in parenthesis. }
\label{Res}
\end{figure}

After obtaining the scattering phase shifts, the partial cross sections for different $\ell^{\pi}$ channels of the $\alpha$-${}^{12}\text{C}$ system can be calculated as \cite{alpha_alpha}
\begin{equation}
\sigma_\ell=\frac{4\pi}{k^2}(2\ell+1)sin^2\delta_\ell(E)
\end{equation}
where $k = \sqrt{2\mu E/\hbar^2}$, $\mu$ is the reduced mass, and $\delta_\ell$ represents the phase shift corresponding to each partial wave.
The partial cross sections for various states as a function of $E_{cm}$, calculated from the obtained phase shift values, are shown in Fig.~\ref{Res}.
From this figure, it is evident that broad and narrow resonances are associated with different orbital angular momentum (\( \ell \)) channels. By analyzing these cross-section plots, the resonance parameters-namely, the resonance energy \( E_r \) and resonance width \( \Gamma_r \), can be extracted. These parameters are obtained by fitting a Gaussian function to the resonance peaks observed in the partial cross-section plots, using Origin Software. The resonance energy \( E_r \) corresponds to the peak of the Gaussian curve, while the resonance width \( \Gamma_r \), defined as the full width at half maximum (FWHM), provides information about the decay rate of the resonant state. The calculated resonance parameters are given in Table \ref{resonance}.
\begin{table}[h!]
\centering
\caption{Calculated resonance energies $E_{r}$ (MeV) and resonance widths for different $\ell^{\pi}$ channels, compared with experimental data \cite{epaps_12C_alpha}.}
\begin{tabular}{|c|c|c|c|c|}
\hline
\textbf{States ($\ell^{\pi}$)} & \textbf{$E_{r}$ (MeV)} &\textbf{$\Gamma_{r}$ (keV)} &\textbf{$E_{r}$ (MeV) \cite{epaps_12C_alpha} } & \textbf{$\Gamma_{r}$ (keV) \cite{epaps_12C_alpha}}\\
\hline
$1^-$ & $2.398 \pm 0.004$ & $410 \pm 11$ & $2.423 \pm 0.011$ & $420 \pm 20$ \\
$2^+$ & $4.355 \pm 0.003$ & $71 \pm 6$ & $4.357 \pm 0.004$ & $73 \pm 5$ \\
$3^-$ & $4.538 \pm 0.008$ & $882 \pm 17$ & $4.440 \pm 0.015$ & $800 \pm 100$ \\
$4^+$ & $3.1728 \pm 0.0001$ & $26 \pm 0.1$ & $3.194 \pm 0.003$ & $27 \pm 3$ \\
\hline
\end{tabular}
\label{resonance}
\end{table}

From Table~\ref{resonance}, we observe that for the $\ell^{\pi}=1^-$ and $\ell^{\pi}=3^-$ states, broad resonances are obtained, and the calculated resonance parameters ($E_r$ and $\Gamma_r$) are in good agreement with the experimental values \cite{epaps_12C_alpha}. The corresponding uncertainties in the calculated resonance parameters are also provided. For the $\ell^{\pi}=2^+$ state, we focus on the second (broad) resonance, as the first (narrow) resonance was not considered in this analysis. The extracted parameters for this broad resonance match well with the experimental data. Similarly, for the $\ell^{\pi}=4^+$ state, we consider only the first resonance, which is a narrow resonance. The calculated energy and width for this state also show excellent agreement with the experimental measurements. Overall, the results demonstrate that the calculated resonance parameters are consistent with experimental observations across different $\ell$ channels.
\section{Conclusions}
In this work, we have constructed the inverse potentials for resonant states of \(\alpha-^{12}\mathrm{C}\) scattering using a machine learning-based optimization technique in conjunction with the variable phase approach. By employing the inverse method, wherein phase shift data are provided as input, we obtain a physically acceptable interaction potential \(V(r)\). The potential \(V(r)\) is modeled as a piecewise combination of three smoothly connected Morse-type functions. A key advantage of this methodology is its ability to naturally incorporate both the nuclear and Coulomb interactions without requiring explicit separate forms for each contribution. Through optimization of the model parameters using a genetic algorithm, the resulting interaction potential successfully reproduces the Coulomb barrier, demonstrating that the electromagnetic effects are inherently captured by the model.\\
Furthermore, the resonance energies and decay widths of the \(\alpha-^{12}\mathrm{C}\) system, determined from the constructed potential, show excellent agreement with experimental data. This confirms that our approach, employing an input potential composed of three smoothly joined Morse functions to solve the phase equation for charged-particle scattering, effectively constructs inverse potentials that accurately reproduce the physical observables. An important outcome of this study is that the extracted Coulomb barrier can be directly utilized in low-energy astrophysical \(S\)-factor calculations, thereby eliminating the need for approximations concerning Coulomb interaction effects. This approach provides a more accurate and robust framework for investigating low-energy nuclear reactions of astrophysical significance. The calculation of the S-factor is currently in progress, and the findings will be reported in future studies.
\\
\textbf{Acknowledgments}
A. Awasthi acknowledges financial support provided by Department of Science and Technology (DST), Government of India vide Grant No. DST/INSPIRE Fellowship/2020/IF200538. 
\\ 
\textbf{Author Declaration} 
The authors declare that they have no conflict of interest.

\end{document}